\documentclass[fleqn]{article}
\DeclareFontFamily{OT1}{rsfs}{}
\DeclareFontShape{OT1}{rsfs}{m}{n}{<5> rsfs5 <7> rsfs7 <10> rsfs10
}{}
\DeclareSymbolFont{mathrsfs}{OT1}{rsfs}{m}{n}
\DeclareSymbolFontAlphabet{\mathrsfs}{mathrsfs}
\begin{document}
\author{R.R. Lompay\thanks{Institute of Electron Physics Nat. Acad. Sci. of
Ukraine, Department of Theory of Elementary Interactions,
Uzhgorod, Ukraine, E-mail: lompay@zk.arbitr.gov.ua}}
\title{Deriving Mathisson - Papapetrou equations from relativistic
pseudomechanics}
\date{}
\maketitle
\newcommand{\dt}{\mathrsfs{D}\theta\,}
\newcommand{\limz}{_{\theta_1}^{\theta_2}}
\newcommand{\intz}{\int\limits\limz\dt }
\newcommand{\intt}{\int\limits\limz d\theta }
\newcommand{\psico}{\overline{\psi}}
\newcommand{\lb}{\left[}
\newcommand{\rb}{\right]}%
\newcommand{\bt}{\begin{array}{c}\mbox{boundary}\\ \mbox{terms}\end{array}}%
\begin{abstract}

It is shown that the equations of motion of a test point particle
with spin in a given gravitational field, so called Mathisson - Papapetrou 
equations, can be derived from Euler - Lagrange
equations of the relativistic pseudomechanics -- relativistic
mechanics, which side by side uses the conventional (commuting)
and Grassmannian (anticommuting) variables. In this approach the
known difficulties of the Mathisson - Papapetrou equations,
namely, the problem of the choice of supplementary conditions and
the problem of higher derivatives are not appear.

\end {abstract}

In ref. \cite{L1} dealing with a description of classical
relativistic point particle with spin was suggested a model based
on action functional
\begin{equation}
W[z,\psi,\psico;\theta_{1,2}]=
\frac{1}{c}\intz\lb-mc^2+\frac{i\hbar c
}{2}(\psico(\theta)D\psi(\theta)-
D\psico(\theta)\psi(\theta))+\frac{\kappa}{2}mc^2\psico(\theta)
\gamma_\alpha\psi(\theta)Dz^\alpha(\theta)\rb.
\end{equation}
Here $c$ is light velocity in vacuum; $\hbar$ -- Planck constant;
$m$ -- bare mass of  the particle; $\kappa$ -- dimensionless
coupling constant; $\theta$ -- an arbitrary parameter which
numerates the points of the world line of the particle and has
length dimension; $z\equiv\{z^\mu\}=(z^0,z^1,z^2,z^3)$ are
pseudocartesian coordinates of the particle in Minkowski space;
$\gamma^\mu\equiv\{\gamma^\mu{}^A{}_B\}$ -- Dirac matrices which
satisfy anticommutation relations
\begin{equation}
\gamma^\mu\gamma^\nu+\gamma^\nu\gamma^\mu=-2\eta^{\mu\nu}I,
\end{equation}
where
\begin{equation}
\{\eta_{\mu\nu}\}=\{\eta^{\mu\nu}\}=\rm{diag}(-1,1,1,1)
\end{equation}
is Minkowski space metrics; $I\equiv\{\delta^A{}_B\}$ -- unit
$4\times4$-matrix; $\psi\equiv\{\psi^A\}$ and
$\psico\equiv\{\psico_B\}=\psi^\dag\gamma^0$ are Dirac spinor and
conjugated to it spinor, accordingly, which are defined on the
world line of the paricle and considered as Grassmannian
variables:
\begin{equation}
\psi^A\psi^B=-\psi^B\psi^A,
\quad\psico_A\psico_B=-\psico_B\psico_A,
\quad\psi^A\psico_B=-\psico_B\psi^A.
\end{equation}

$\dt\equiv[-\eta_{\alpha\beta}
z'^\alpha(\theta)z'^\beta(\theta)]^{1/2}d\theta$\, and
$D\equiv[-\eta_{\alpha\beta}
z'^\alpha(\theta)z'^\beta(\theta)]^{-1/2}d/d\theta$\, are,
accordingly, reparametrization-invariant (RI) measure and
RI-derivative –- the objects which allow to fulfill  the
calculations not depending on the special choice of the parameter
$\theta$ (see ref. \cite{L2}); $z'^\alpha(\theta)\equiv
dz^\alpha(\theta)/d\theta$.

The model (1) correctly describes the kinematical connection
between orbital and spin parts of the total 4-angular momentum
$J_{\rho\sigma}=M_{\rho\sigma}+S_{\rho\sigma}$ and gives also
true results for Poisson brackets for dynamical variables in
corresponding Hamilton formalism.

In this work we show that in the case of interaction such
particle with an external gravitational field the equation of
motion which follow from corresponding generalization of the
action functional (1), leads to Mathisson – Papapetrou equation
(MPE) \cite{M,P}.

We introduce an interaction of the particle with gravitational
field with minimal coupling by covariantization of action
functional (1). Metrics (3) and $\gamma$-matrices (2) of Minkowski
space, which we shall denote now as $\eta_{(\alpha)(\beta)}$ and
$\gamma^{(\alpha)}$, we replace by metrics $g_{\mu\nu}(x)$ and
$\gamma$-matrices $\gamma^\mu(x)$ of an arbitrary pseudorimannian
space. And RI-deriative $D$ we replace by absolute RI-deriative
$\nabla$. Then the action functional (1) take the form
\begin{equation}
\begin{array}{l}
W[z,\psi,\psico,g;\theta_{1,2}]=\\
\quad=\frac{1}{c}\intz\lb-mc^2+\frac{i\hbar c
}{2}(\psico(\theta)\nabla\psi(\theta)-
\nabla\psico(\theta)\psi(\theta))+\frac{\kappa}{2}mc^2\psico(\theta)
\gamma_\alpha(z(\theta))\psi(\theta)Dz^\alpha(\theta)\rb\\
\quad\equiv W_1+W_2+W_3,
\end{array}
\end{equation}
where now $\dt=[-g_{\alpha\beta}(z(\theta))
z'^\alpha(\theta)z'^\beta(\theta)]^{1/2}d\theta$;\,
$D=[-g_{\alpha\beta}(z(\theta))
z'^\alpha(\theta)z'^\beta(\theta)]^{-1/2}d/d\theta$;\,
$\gamma^\mu(x)\equiv\gamma^{(\alpha)}e_{(\alpha)}{}^\mu(x)$ and
$e_{(\alpha)}{}^\mu(x)$ form a fierbein:
\begin{equation}
\eta^{(\alpha)(\beta)}e_{(\alpha)}{}^\mu(x)e_{(\beta)}{}^\nu(x)=g^{\mu\nu}(x);\quad
g_{\mu\nu}(x)e_{(\alpha)}{}^\mu(x)e_{(\beta)}{}^\nu(x)=\eta_{(\alpha)(\beta)},
\end{equation}
so that
\begin{equation}
\gamma^\mu(x)\gamma^\nu(x)+\gamma^\nu(x)\gamma^\mu(x)=-2g^{\mu\nu}(x)I.
\end{equation}
Absolute RI-derivative $\nabla$ acting on the spinors is defined
as
\begin{equation}
\begin{array}{l}
\nabla\psi(\theta)\equiv D\psi(\theta)+
\Omega_\nu(z(\theta))\psi(\theta)Dz^\nu(x);\\
\nabla\psico(\theta)\equiv D\psico(\theta)-
\psico(\theta)\Omega_\nu(z(\theta))Dz^\nu(x),
\end{array}
\end{equation}
where $\Omega_\nu(x)$ are spin connection coeffitients:
\begin{equation}
\Omega_\nu(x)=\frac{1}{2!}s^{(\rho)(\sigma)}\omega_{(\rho)(\sigma)(\alpha)}(x)
e^{(\alpha)}{}_\nu(x);\quad\omega_{(\rho)(\sigma)(\alpha)}(x)\equiv
e_{(\rho)}{}^\mu(x)(\nabla_\lambda
e_{(\sigma)\mu}(x))e_{(\alpha)}{}^\lambda(x).
\end{equation}
And the action on the vector is defined as
\begin{equation}
\nabla Dz^\mu(\theta)\equiv D^2z^\mu(\theta)+
\Gamma^\mu{}_{\lambda\nu}(z(\theta))Dz^\lambda(\theta)
Dz^\nu(\theta),
\end{equation}
where $\Gamma^\mu{}_{\lambda\nu}(x)$ are Kristoffel symbols:
\begin{equation}
\Gamma^\mu{}_{\lambda\nu}(x)\equiv
g^{\mu\alpha}(x)\frac{1}{2}(\partial_\lambda g_{\alpha\nu}(x)+
\partial_\nu g_{\alpha\lambda}(x)- \partial_\alpha g_{\lambda\nu}(x)).
\end{equation}

Let as find the variations of the action functional (8), caused
by varying of quantities $z(\theta),\psi(\theta),\psico(\theta)$:
\begin{equation}
\delta W[z,\psi,\psico,g;\theta_{1,2}]\equiv W[z+\delta z
,\psi+\delta\psi,\psico+\delta\psico,g;\theta_{1,2}]-
W[z,\psi,\psico,g;\theta_{1,2}]=\delta_z W+\delta_\psi
W+\delta_{\psico} W.
\end{equation}
\begin{equation}
\delta_z W[z,\psi,\psico,g;\theta_{1,2}]\equiv W[z+\delta
z,\psi,\psico,g;\theta_{1,2}]-W[z,\psi,\psico,g;\theta_{1,2}]=
\delta_z W_1+\delta_z W_2+\delta_z W_3;
\end{equation}
\begin{equation}
\begin{array}{l}
\delta_z W_1=-mc\int\limits\limz\delta(\dt)=-mc\intt\,
\delta_z\left(\sqrt{-g_{\alpha\beta}(z) z'^\alpha
z'^\beta}\right)\\
\quad=\frac{mc}{2}\intt\, [-g_{\rho\sigma}(z)z'^\rho
z'^\sigma]^{-1/2}\left(\partial_\mu g_{\alpha\beta}\delta z^\mu
z'^\alpha z'^\beta+2g_{\alpha\mu}z'^\alpha(\delta
z^\mu)'\right)\\
\quad=\frac{mc}{2}\intz\lb\partial_\mu g_{\alpha\beta}Dz^\alpha
Dz^\beta\delta z^\mu+2g_{\alpha\mu}Dz^\alpha D(\delta z^\mu)\rb\\
\quad=\frac{mc}{2}\intz\lb\partial_\mu g_{\alpha\beta}Dz^\alpha
Dz^\beta-2D(g_{\alpha\mu}Dz^\alpha)\rb\delta z^\mu+\bt\\
\quad=-mc\intz\lb g_{\alpha\mu}D^2
z^\alpha+\frac{1}{2}(\partial_\alpha g_{\beta\mu}+ \partial_\beta
g_{\alpha\mu}-\partial_\mu g_{\alpha\beta})Dz^\alpha
Dz^\beta\rb\delta z^\mu+\bt\\
\quad=\intz\lb-\nabla(mcg_{\lambda\mu}Dz^\lambda)\rb\delta
z^\mu+\bt;
\end{array}
\end{equation}
\begin{equation}
\begin{array}{l}
\delta_z
W_2=\frac{i\hbar}{2}\intt\lb\psico\delta_z(\psi'+\Omega_\alpha(z)\psi
z'^\alpha)-\delta_z(\psico'-\psico\Omega_\alpha(z)
z'^\alpha)\psi\rb=i\hbar\intt\psico\delta_z(\Omega_\alpha(z)
z'^\alpha))\psi\\
\quad=i\hbar\intt\lb\psico\partial_\mu\Omega_\alpha(z)\psi
z'^\alpha\delta z^\mu+\psico\Omega_\mu(z)\psi(\delta z^\mu)'\rb\\
\quad=i\hbar\intz\lb\psico\partial_\mu\Omega_\alpha(z)\psi
Dz^\alpha\delta z^\mu+\psico\Omega_\mu(z)\psi D(\delta z^\mu)\rb\\
\quad=i\hbar\intz\lb\psico(\partial_\mu\Omega_\alpha)\psi
Dz^\alpha-
D(\psico\Omega_\mu\psi)\rb\delta z^\mu+\bt\\
\quad=i\hbar\intz\lb\psico(\partial_\mu\Omega_\alpha-
\partial_\alpha\Omega_\mu)\psi Dz^\alpha- D\psico\Omega_\mu\psi-\psico\Omega_\mu
D\psi\rb\delta z^\mu+\bt\\
\quad=i\hbar\intz\lb\psico(\partial_\mu\Omega_\alpha-
\partial_\alpha\Omega_\mu+\Omega_\mu\Omega_\alpha-\Omega_\alpha\Omega_\mu)\psi
Dz^\alpha-\right.\\ \quad\left.-(D\psico-\psico\Omega_\alpha
Dz^\alpha)\Omega_\mu\psi-\psico\Omega_\mu(D\psi+\psi\Omega_\alpha
Dz^\alpha)\rb\delta z^\mu+\bt\\
\quad=\intz\lb\frac{1}{2}(\hbar\psico
s^{\rho\sigma}\psi)R_{\rho\sigma\mu\alpha}
Dz^\alpha-i\hbar(\nabla\psico\Omega_\mu\psi+\psico\Omega_\mu\nabla\psi)\rb\delta
z^\mu+\bt;
\end{array}
\end{equation}
\begin{equation}
\begin{array}{l}
\delta_z
W_3=\frac{\kappa}{2}mc\intt\psico\delta_z(\gamma_\alpha(z)
z'^\alpha))\psi=
\frac{\kappa}{2}mc\intz\lb\psico(\partial_\mu\gamma_\alpha)\psi
Dz^\alpha-
D(\psico\gamma_\mu\psi)\rb\delta z^\mu+\bt\\
\quad=\frac{\kappa}{2}\intz\lb\psico(\partial_\mu\gamma_\alpha-
\Gamma^\beta{}_{\alpha\mu}\gamma_\beta)\psi Dz^\alpha-
(D(\psico\gamma_\mu\psi)-\Gamma^\nu{}_{\mu\alpha}(\psico\gamma_\nu\psi)
Dz^\alpha)\rb\delta z^\mu+\bt\\
\quad=\intz\lb\frac{\kappa}{2}mc\psico
(\gamma_\alpha\Omega_\mu-\Omega_\mu\gamma_\alpha)\psi
Dz^\alpha-\nabla(\frac{\kappa}{2}mc\psico\gamma_\mu\psi)\rb\delta
z^\mu+\bt.
\end{array}
\end{equation}
In obtaining the last equalities in (15) and (16) we used the
identities
\begin{equation}
\partial_\mu\Omega_\alpha-\partial_\alpha\Omega_\mu+
\Omega_\mu\Omega_\alpha-\Omega_\alpha\Omega_\mu\equiv
\frac{i}{2!}s^{\rho\sigma}R_{\rho\sigma\mu\alpha},\quad
s^{\rho\sigma}\equiv\frac{i}{2}(\gamma^\rho \gamma^\sigma-
\gamma^\sigma \gamma^\rho).
\end{equation}
\begin{equation}
\nabla_\mu\gamma_\alpha\equiv \partial_\mu\gamma_\alpha-
\Gamma^\beta{}_{\alpha\mu}\gamma_\beta+ \Omega_\mu\gamma_\alpha-
\gamma_\alpha\Omega_\mu\equiv 0.
\end{equation}
Furthermore,
\begin{equation}
\delta_\psi W[z,\psi,\psico,g;\theta_{1,2}]\equiv W[z,\psi+\delta
\psi,\psico,g;\theta_{1,2}]-W[z,\psi,\psico,g;\theta_{1,2}]=
\delta_\psi W_2+\delta_\psi W_3;
\end{equation}
\begin{equation}
\begin{array}{l}
\delta_\psi
W_2=\frac{i\hbar}{2}\intz\lb\psico\delta_\psi(D\psi+\Omega_\alpha\psi
Dz^\alpha)-\nabla\psico\delta\psi\rb=
\frac{i\hbar}{2}\intz\lb\psico
D(\delta\psi)+\psico\Omega_\alpha\delta\psi
Dz^\alpha-\nabla\psico\delta\psi\rb\\
\quad=\frac{i\hbar}{2}\intz\lb-(D\psico-\psico\Omega_\alpha
Dz^\alpha)-\nabla\psico\delta\psi\rb\delta\psi+\bt\\
\quad=\intz\lb-i\hbar\nabla\psico\rb\delta\psi+\bt;
\end{array}
\end{equation}
\begin{equation}
\begin{array}{l}
\delta_\psi W_3=\intz\lb\frac{\kappa}{2}mc\psico\gamma_\alpha
Dz^\alpha\rb\delta\psi.
\end{array}
\end{equation}
Analogously
\begin{equation}
\delta_{\psico} W[z,\psi,\psico,g;\theta_{1,2}]\equiv W[z,\psi,
\psico+\delta{\psico},g;\theta_{1,2}]-W[z,\psi,\psico,g;\theta_{1,2}]=
\delta_{\psico} W_2+\delta_{\psico} W_3;
\end{equation}
\begin{equation}
\begin{array}{l}
\delta_{\psico} W_2= \intz\delta\psico\lb i\hbar\nabla\psi\rb+\bt;
\end{array}
\end{equation}
\begin{equation}
\begin{array}{l}
\delta_{\psico}
W_3=\intz\delta\psico\lb\frac{\kappa}{2}mc\gamma_\alpha\psi
Dz^\alpha\rb.
\end{array}
\end{equation}
Taking, as usually, that on the boundary of integration
$\theta=\theta_{1,2}$, the variations vanish, $\delta
z_{1,2}=\delta \psi_{1,2}=\delta\psico_{1,2}=0$, using the
formulae (12) – (16), (19) – (24), we obtain the following
equations of the motion
\begin{equation}
\frac{\Delta W}{\Delta z^\mu}=-\nabla\lb
mcg_{\lambda\mu}(Dz^\lambda+\frac{\kappa}{2}\psico\gamma^\lambda\psi)\rb+\frac{1}{2!}(\hbar\psico
s^{\rho\sigma}\psi)R_{\rho\sigma\mu\nu}Dz^\nu-
\end{equation}
$$
\quad-(i\hbar\nabla\psico-\frac{\kappa}{2}mc\psico\gamma_\alpha
Dz^\alpha)\Omega_\mu\psi-\psico\Omega_\mu(i\hbar\nabla\psi+\frac{\kappa}{2}mc\gamma_\alpha\psi
Dz^\alpha)=0;
$$
\begin{equation}
\frac{\Delta^R
W}{\Delta\psi}=-(i\hbar\nabla\psico-\frac{\kappa}{2}mc\psico\gamma_\alpha
Dz^\alpha)=0;
\end{equation}
\begin{equation}
\frac{\Delta^L
W}{\Delta\psico}=i\hbar\nabla\psi+\frac{\kappa}{2}mc\gamma_\alpha\psi
Dz^\alpha=0.
\end{equation}
Substituting the eqs. (26), (27) into eq. (25), we reduce the
last system to the form
\begin{equation}
\left\{\begin{array}{l} \nabla\lb
mc(Dz^\mu+\frac{\kappa}{2}\psico\gamma^\mu\psi)\rb=
\frac{1}{2!}R^\mu{}_{\nu\rho\sigma}Dz^\nu(\hbar\psico
s^{\rho\sigma}\psi);\\
\nabla\psi=\frac{i}{2}\kappa(\frac{mc}{\hbar})\gamma_\alpha\psi
Dz^\alpha;\\
\nabla\psico=-\frac{i}{2}\kappa(\frac{mc}{\hbar})\psico\gamma_\alpha
Dz^\alpha.
\end{array}\right.
\end{equation}
Using the notation
\begin{equation}
P^\mu\equiv
m\left(Dz^\mu+\frac{\kappa}{2}\psico\gamma^\mu(z)\psi\right),
\quad S_{\rho\sigma}\equiv\psico s_{\rho\sigma}(z)\psi
\end{equation}
for the expressions which are the general covariant
generalization of the expressions for 4-momentum and 4-spin of
the particles in pseudoeuclidian space-time \cite{L1}, we can
write the first eq. in (28) also in the form
\begin{equation}
\nabla P^\mu=\frac{1}{2!}R^\mu{}_{\nu\rho\sigma}Dz^\nu
S^{\rho\sigma}.
\end{equation}
With the help of eqs. (28),(29) it is easy to obtain the equations
\begin{equation}
\nabla S_{\rho\sigma}=-(Dz_\rho P_\sigma-Dz_\sigma P_\rho),
\end{equation}
As consequences of these equation one obtain the equations
\begin{equation}
\nabla S_{\rho\sigma}+ Dz_\rho Dz^\nu S_{\nu\sigma}+ Dz_\sigma
Dz^\nu S_{\rho\nu}=0,
\end{equation}
and
\begin{equation}
Dz^\nu\nabla S_{\nu\mu}=(\delta^\nu{}_\mu+Dz^\nu Dz_\mu)P_\nu.
\end{equation}
Using the eqs. (33), one can represent the 4-momentum in the form
\begin{equation}
P^\mu=-Dz^\mu Dz_\nu P^\nu+(\delta^\nu{}_\mu+Dz^\nu
Dz_\mu)P_\nu=MDz^\mu+Dz^\nu\nabla S_\nu{}^\mu, \quad
M\equiv-Dz_\nu P^\nu
\end{equation}
Substituting the decomposition (34) into the eq. (30), one obtain
\begin{equation}
\nabla(MDz^\mu+Dz^\nu\nabla
S_\nu{}^\mu)=\frac{1}{2!}R^\mu{}_{\nu\rho\sigma}Dz^\nu
S^{\rho\sigma}.
\end{equation}
The system of eqs. (32), (35)
\begin{equation}
\left\{\begin{array}{l} \nabla S_{\rho\sigma}+ Dz_\rho Dz^\nu
S_{\nu\sigma}+ Dz_\sigma Dz^\nu S_{\rho\nu}=0;\\
\nabla(M Dz^\mu+Dz^\nu\nabla
S_\nu{}^\mu)=\frac{1}{2!}R^\mu{}_{\nu\rho\sigma}Dz^\nu
S^{\rho\sigma}
\end{array}\right.
\end{equation}
at first were obtain in papers \cite{M,P} and now is known as
Mathisson – Papapetrou equations. The first equation of the system
is called the equation of motion of the spin and second one --
the equation of motion of the particle in space.

In original papers \cite{M,P} these equations were derived
halfphenomenologically by integrating of multiple 3-momentum of a
matter distributing in space with posterior reducing it to a
point and covariantization the obtaining results. In our
consideration the MPE are obtained as exact consequences of the
variation Euler - Lagrange equations for abhoc point generally
covariant physical system. This approach is automatically free
from the well known difficulties MPE, namely.

The first is the problem of choice of supplementary conditions.
If the eqs. (36) are considered as the closed system for
determining $z^\mu(\theta)$ and $S_{\rho\sigma}(\theta)$, then it
is easy to see, that the number of equations is less then the
number of unknown functions. The first eq. (36) determines not
the total $\nabla S_{\rho\sigma}$, but only its space projection
\begin{equation}
h^\mu_\rho h^\nu_\sigma\nabla
S_{\mu\nu}\equiv(\delta^\mu{}_\rho+Dz^\mu Dz_\rho)
(\delta^\nu{}_\sigma+Dz^\nu Dz_\sigma)\nabla S_{\mu\nu}= \nabla
S_{\rho\sigma}+ Dz_\rho Dz^\nu S_{\nu\sigma}+ Dz_\sigma Dz^\nu
S_{\rho\nu}.
\end{equation}
So, the system MPE is not fully determined. For full
determination one puts on $S_{\rho\sigma}$ some supplementary
conditions, such as
\begin{equation}
\mbox{Corinaldesi-Papapetrou-Pirani condition \cite{CP,Pi}} \qquad
S^{\mu\nu}Dz_\nu=0;
\end{equation}
\begin{equation}
\mbox{Tulczyjew condition \cite{T}}\qquad S^{\mu\nu}P_\nu=0
\end{equation}
and some others. The problem is that there is not essential
reason to prefer one or other supplementary condition. In our
approach based on action functional (5) this problem doesn't
arises. Indeed, the eqs. (36) are only a sequence of the basic
system of eqs. (28) which fully determines all fundamental
variables $z(\theta),\psi(\theta),\psico(\theta)$ when initial
data $z_0\equiv
z(\theta_0),\psi_0\equiv\psi(\theta_0),\psico_0\equiv\psico(\theta_0)$
are given. By known fundamental variables all other dynamical
variables are determined unique. So, for example,  the 4-spin is
given by the second formula in (29).

The second is the problem of higher derivatives. As it is easy to
see, the left-hand side of the second eq. in (36) involves the
second derivative of $S_{\rho\sigma}$, which is not determined,
since, as were noted, the first eq. in (36) fixes only the space
part of $\nabla S_{\rho\sigma}$. When the supplementary condition
(38) is imposed, the expression in parentheses in the second eq.
in (36) can be rewritten as $(MDz^\mu-\nabla(Dz^\nu)
S_\nu{}^\mu)$. Then the left-hand side of the second eq. in (36)
will include the third derivative of coordinates of the particle,
but this contradicts to traditional form of mechanics. In our
approach this problem is not arising, too. In accordance with the
equations given above, the expression in parentheses in second
eq. in (36) is equal to
$m(Dz^\mu+\frac{\kappa}{2}\psico\gamma^\mu\psi)$, and higher
derivatives are absent.
\bibliographystyle{osa}
\bibliography{Lompay02}
\end{document}